\documentclass{article}

\usepackage{graphicx}%
\usepackage{multirow}%
\usepackage{amsmath,amssymb,amsfonts}%
\usepackage{amsthm}%
\usepackage{mathrsfs}%
\usepackage[title]{appendix}%
\usepackage{textcomp}%
\usepackage{manyfoot}%
\usepackage{booktabs}%
\usepackage{algorithm}%
\usepackage{algorithmicx}%
\usepackage{algpseudocode}%
\usepackage{listings}%
\usepackage{bm}

\usepackage{chemformula}
\usepackage{siunitx}
\usepackage{csquotes}
\usepackage{comment}
\usepackage{lineno}

\usepackage{color}
\usepackage{tabularray}
\usepackage{array}
\usepackage{multirow}
\usepackage{tabulary}

\usepackage{lipsum}

\title{Revealing the Altermagnetism in Hematite via XMCD Imaging and Anomalous Hall Electrical Transport} 
\author{Edgar Galindez-Ruales\\
Institute of Physics,\\ Johannes Gutenberg University Mainz, \\Staudingerweg 7, Mainz, 55128, Germany.\\
\and
Rafael Gonzalez-Hernandez\\
Grupo de Investigación en Física Aplicada,\\ Departamento de Física,
Universidad del Norte, Barranquilla, Colombia.\\
Fachbereich Physik, \\Universität Konstanz,\\ Universitätsstr. 10, Konstanz, D-78457, Germany.\\
\and
Christin Schmitt\\
Institute of Physics,\\ Johannes Gutenberg University Mainz, \\Staudingerweg 7, Mainz, 55128, Germany.
\\
\and
Shubhankar Das\\
Institute of Physics,\\ Johannes Gutenberg University Mainz, \\Staudingerweg 7, Mainz, 55128, Germany.
\\
\and
Felix Fuhrmann\\
Institute of Physics, \\Johannes Gutenberg University Mainz, \\Staudingerweg 7, Mainz, 55128, Germany.
\\
\and
Andrew Ross\\
Institute of Physics,\\ Johannes Gutenberg University Mainz,\\ Staudingerweg 7, Mainz, 55128, Germany.
\\
\and
Evangelos Golias\\
MAX IV Laboratory, \\
Fotongatan 8, 22484 Lund, Sweeden.
\\
\and
Akashdeep Akashdeep\\
Institute of Physics, \\Johannes Gutenberg University Mainz,\\ Staudingerweg 7, Mainz, 55128, Germany.
\\
\and
Laura Lünenbürger\\
Institute of Physics,\\ Johannes Gutenberg University Mainz,\\ Staudingerweg 7, Mainz, 55128, Germany.
\\
\and
Eunchong Baek\\
Institute of Physics,\\ Johannes Gutenberg University Mainz,\\ Staudingerweg 7, Mainz, 55128, Germany.\\
Department of Physics and Chemistry,\\
DGIST, Daegu 42988, Republic of Korea.\\
\\
\and
Wanting Yang\\
Materials Genome Institute,\\ Institute of Quantum Science and Technology,\\ International Center for Quantum and Molecular Structures,\\ Shanghai University, \\99 Shangda Road, Shanghai, 200444, China.\\
\and
Libor Šmejkal\\
Institute of Physics,\\ Johannes Gutenberg University Mainz,\\ Staudingerweg 7, Mainz, 55128, Germany.
\\
\and
Venkata Krishna\\
Institute of Physics,\\ Johannes Gutenberg University Mainz,\\ Staudingerweg 7, Mainz, 55128, Germany.
\\
\and
Rodrigo Jaeschke\\
Institute of Physics,\\ Johannes Gutenberg University Mainz,\\ Staudingerweg 7, Mainz, 55128, Germany.
\\
\and
Jairo Sinova\\
Institute of Physics, \\Johannes Gutenberg University Mainz, \\Staudingerweg 7, Mainz, 55128, Germany.
\\
\and
Avner Rothschild\\
Department of Materials Science and Engineering,\\
Technion-Israel Institute of Technology, \\
Haifa 32000, Israel.\\
\\
\and
Chun-Yeol You\\
Department of Physics and Chemistry,\\
DGIST, Daegu 42988, Republic of Korea.\\
\\
\and
Gerhard Jakob\\
Institute of Physics, \\Johannes Gutenberg University Mainz, \\Staudingerweg 7, Mainz, 55128, Germany.
\\
\and
Mathias Kläui\\
Institute of Physics,\\ Johannes Gutenberg University Mainz,\\ Staudingerweg 7, Mainz, 55128, Germany.\\
Center for Quantum Spintronics,\\ Norwegian University of Science and Technology,\\ Høgskoleringen 5, Trondheim, 7034, Norway.\\
\texttt{klaeui@uni-mainz.de}
}
\begin{document}
\maketitle

\begin{abstract}
Altermagnets are a class of magnetic materials that exhibit unconventional transport properties, such as an anomalous Hall effect, despite having compensated sublattice magnetic moments. In this study, we report fundamental experimental evidence of the altermagnetic nature of hematite ($\alpha$-\ch{Fe2O3}), combining electrical transport with advanced XPEEM imaging with linear and circular dichroism contrast. Our measurements directly visualize the Néel vector's coupling to the crystal orientation, confirming hematite's altermagnetic order and its symmetry-driven transport behavior. \color{black} The\color{black} transport measurements reveal an anisotropic AHE with a pronounced crystal orientation dependence, including a sign inversion for specific Néel vector alignments. Supported by first-principles theoretical calculations, we explain how the interplay between collinear spin and crystal symmetry breaking drives the observed anomalous Hall effect. These findings establish hematite as an altermagnet, paving the way for experimental identification of altermagnetic materials and their integration into spintronic technologies.
\end{abstract} 






\maketitle

\section{Introduction}\label{sec1}
Altermagnets are an emerging class of magnetic materials characterized by compensated d, g, or i-wave spin order that breaks time-reversal symmetry in both direct and reciprocal space, enabling unconventional transport phenomena such as a nonzero anomalous Hall effect (AHE) without net magnetization \cite{LiborCHE, PRXLibor2022,bai_nonlinear_2025,urata_high_2024,betancur, Reichlova2024, RuO, Bai_RuO2, Mn5Si3_Biniskos}. \color{black} Altermagnets can show unconventional transport phenomena (such as noncollinear spin current without spin orbit coupling (SOC) \cite{ma_multifunctional_2021}) and AHE without net magnetization, when magnetic point group symmetry is compatible with ferromagnetism \cite{Anomalus_antiferro, gonzalez-hernandez_efficient_2021} \color{black}. Unlike conventional antiferromagnets, where the AHE is forbidden due to strict symmetry constraints, altermagnets exhibit symmetry-driven transport properties arising from the interplay between collinear magnetic order and crystal symmetry \cite{RuO, ChirialHall}, where AHE can also appear in noncollinear and noncoplanar compensated magnets \cite{Ch8, Ch9}. These properties make collinear altermagnets uniquely suited for applications where symmetry and transport phenomena can be exploited, such as spintronics technologies \cite{CHEFeMn, ChirialHall, Ch12, Ch13}. Despite these promising properties, experimental evidence of altermagnetic behavior remains limited, 
with reciprocal space \cite{arpes_sonka, MnTe, PRL_2024_Lee} and real space \cite{Amin2024} evidence in \ch{MnTe} and \ch{CrSb}, and indirect magnetic circular dichroism (MCD), AHE, and spin torque evidence in \ch{MnTe} \cite{betancur}, \ch{RuO2} \cite{RuO, Bose2022, arpes_RuO} and \ch{Mn5Si3} \cite{Reichlova2024}. However, analogous imaging and symmetry-resolved studies in hematite have yet to be reported.

Hematite ($\alpha$-\ch{Fe2O3}), a collinear magnet, has been theoretically classified as an altermagnet \cite{LiborCHE}  (see Table \ref{table:teo}). At the Morin temperature \cite{MorinPaper}, it transitions between a canted weak ferromagnetic (WF) state at high temperatures (above the Morin transition, $T^{bulk}_M=259$ K) and a perfectly compensated collinear phase at lower temperatures \cite{MorinPaper}. These distinct relativistic orders make hematite an intriguing platform for exploring spin transport phenomena, particularly symmetry-driven effects \cite{training, TD_Uli}. While its magnetic anisotropies and Morin transition \color{black}, including the thickness dependence, \color{black} have been extensively studied \cite{AntiferroSpint, Meer_Review, nature590}, its transport properties linked to altermagnetic symmetry remain largely unexplored. Theoretical studies predict that hematite should exhibit a nonzero AHE with strong angular dependence, driven by its crystal symmetry \cite{LiborCHE, CHEFeMn}. However, experimental verification has been hindered by hematite's insulating nature and the difficulty of isolating symmetry-specific transport signatures.

\begin{table}
\centering
\caption{\color{black} The different magnetic phases of altermagnetic hematite below the Néel temperature range (bulk values). The Hall pseudovector is allowed for the collinear ordering with weak ferromagnetism (canted WF), and the ferrimagnetic secondary ordering, but not for the perfectly compensated phase. Due to its real and reciprocal space symmetries, hematite is an altermagnet in all the magnetic phases. Still, only when the magnetic point group is compatible with ferromagnetism, a Hall pseudovector is allowed. The direction of $(\sigma_\text{yz},   \sigma_\text{xz}, \sigma_\text{xy} )$ depends on the orientation of the Néel vector, but it is not orthogonal to it or parallel to the magnetization.  \color{black} Here, $SO(2)$ is a continuous rotation around the axes of spins, $C_2T$ is spin rotation around the axes perpendicular to the spins, and $T$ is time-reversal symmetry.}
\label{table:teo}
\begin{tabular}{|c|c|c|c|c|c|c|}
\hline
\textbf{\begin{tabular}[c]{@{}c@{}}Primary\\ Magnetic\\ Order\end{tabular}} & \textbf{\begin{tabular}[c]{@{}c@{}}Spin\\ Space\\ Group\end{tabular}} & \textbf{\begin{tabular}[c]{@{}c@{}}Spin\\ Point\\ Group\end{tabular}}     & \textbf{\begin{tabular}[c]{@{}c@{}}Spin\\ Orientation\\   ($T$ (K),\\ Secondary \\ Order)\end{tabular}}              & \textbf{\begin{tabular}[c]{@{}c@{}}Magnetic\\ Space\\ Group\end{tabular}} & \textbf{\begin{tabular}[c]{@{}c@{}}Magnetic\\ Point\\ Group\end{tabular}} & \textbf{\begin{tabular}[c]{@{}c@{}}Hall\\ Vector\\ \color{black} (Néel dep.) \color{black}\end{tabular}} \\ \hline
\textbf{}                                                                   &                                                                       &                                                                           & \begin{tabular}[c]{@{}c@{}}$(100)$ \\ ( $T>259$, \\ canted WF)\end{tabular}                       & $C_2’/c’$                                                                    & $2’/m’$                                                                     & $(\sigma_\text{yz},   \sigma_\text{xz} , \sigma_\text{xy} )$                                                \\ 
\textbf{AM}                                                                 & \begin{tabular}[c]{@{}c@{}}$1-32c$\\ $\times SO(2)$\\ $\times C_2T$\end{tabular}         & \begin{tabular}[c]{@{}c@{}}$1-32m$\\ $\times SO(2)$\\ $\times C_2T$\\ (g-wave)\end{tabular} & \begin{tabular}[c]{@{}c@{}}$(001)$\\ ($10<T<259$, \\ perfectly\\ compensated)\end{tabular} & R-3c                                                                      & -31m                                                                      & (0,0,0)                                                        \\
\textbf{}                                                                   &                                                                       &                                                                           & \begin{tabular}[c]{@{}c@{}}Tilted\\ ($T<$10, \\ ferrimagnet)\end{tabular}                            & P-1                                                                       & -1                                                                        & $(\sigma_\text{yz},   \sigma_\text{xz} , \sigma_\text{xy} )$  \\
\hline
\end{tabular}
\end{table}

Low-level doping offers a solution for transport measurements, inducing electronic conductivity while preserving hematite's compensated magnetic order. Doping concentrations below 2\% maintain a stable carrier concentration and minimize lattice stress, which can shift the Morin transition to slightly higher temperatures \cite{ZnHematite, SnHematite, Sihematite}. For heavily doped samples, the Hall voltage scales linearly with the applied field, following conventional behavior \color{black} and distorting the crystal structure  \cite{12Si}. On the other hand, \color{black} for too low-doped samples, the Hall coefficient was undetectable in previous studies \cite{54Si,55Si}. At doping levels below 2\%, the dopants are reported to be fully ionized at all temperatures, ensuring a stable carrier concentration with minimal impact on magnetism \cite{Sihematite, 13Si}. By carefully controlling the doping concentration, the Fermi level can be shifted by a few tenths of an electronvolt \cite{fermilevelpinning, fermilimits, Impurityband}, while maintaining hematite's collinear antiferromagnetic order of the sub-lattices \color{black} and the altermagnetic crystal structure. \color{black}

Complementarily, hematite with different conductivities can be studied using X-ray photoemission electron microscopy (XPEEM) \cite{RcutReference, SMRthinfilms, ZnHematite}; this is a powerful technique for probing materials' spin and orbital properties with nanoscale spatial resolution. In the altermagnetic regime, where the AHE arises from the unique relativistic symmetry properties of the Néel vector and nonrelativistic altermagnetic band spin-splitting, XPEEM provides direct insights into the underlying spin configurations and domain structure. By combining X-ray magnetic linear dichroism (XMLD) and X-ray magnetic circular dichroism (XMCD), XPEEM enables a complete Néel vector mapping, capturing its directionality and the associated symmetry-breaking effects \cite{Amin2024}. \color{black} This imaging capability is indispensable, as it directly confirms that a 180° reversal of the Néel vector results in an inversion of the altermagnetic Hall vector—information not easily accessible through transport measurements alone. However, due to technical constraints, in-situ high-field Hall transport measurements are not possible within the PEEM setup, as strong magnetic fields are incompatible with electron-optical imaging conditions. Consequently, transport measurements and PEEM imaging are complementary techniques, together providing a comprehensive and conclusive picture of the altermagnetic nature of hematite.\color{black}

In this study, we combine \color{black} angle-dependent \color{black} transport measurements with XPEEM to reveal the altermagnetic properties of hematite. Using 1\% \ch{Ti} doping, we realize electronic conduction and investigate the AHE, revealing a pronounced angular dependence, including periodic sign inversions. We utilize first-principles calculations to elucidate the angular dependence of the anomalous Hall effect (AHE) and its absence in certain orientations of the Néel vector. Finally, the combination of XMLD-PEEM and XMCD-PEEM imaging provides direct visualization of the absolute Néel vector orientation, confirming the altermagnetic symmetry-driven origins of the transport properties. 

\section{Results and Discussion}\label{sec2}
\begin{figure}[h]
\centering
\includegraphics[width=0.9\textwidth]{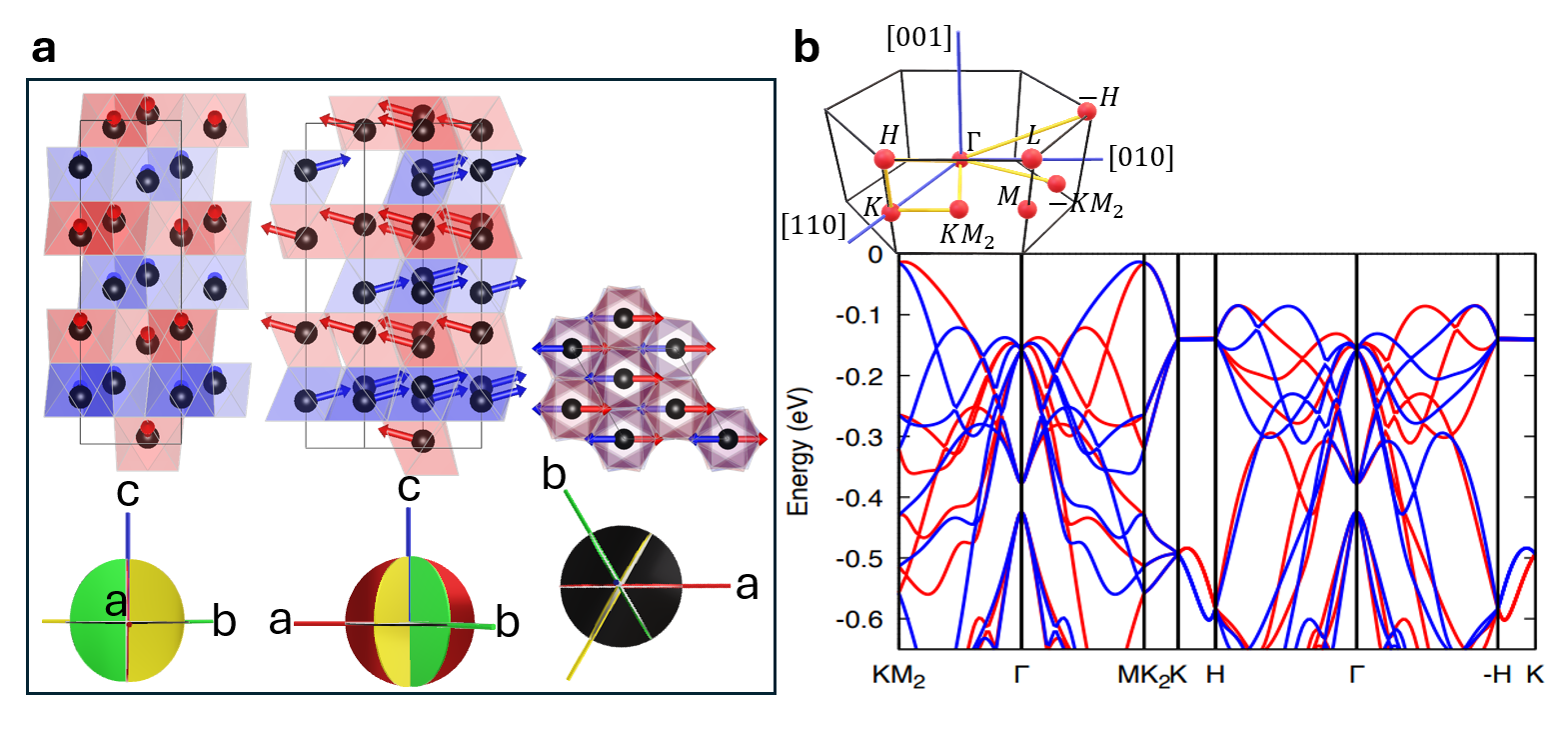}
\caption{\textbf{a} Crystal structure of hematite in the conventional hexagonal unit cell with the oxygen octahedra \color{black} projected along [100], [110], and [001]. The two sublattices (red and blue octahedra) have an arrow that represents the spin orientation of the \ch{Fe} atoms with the canted moment (exaggerated) locked in the \textbf{a}-\textbf{c} plane (red plane), equivalent to the \textbf{b}-\textbf{c} (green) and \textbf{b'}-\textbf{c} (yellow) planes. Above the spinflop, with a field along [0001], the canted moment is also along \textbf{c}, but at zero magnetic field, it lies in the $x$-$y$ plane (black plane). The non-centrosymmetric positions of the oxygen atoms break $T\tau$, while the $PT$ is broken due to the centrosymmetric position of the \ch{Fe} atoms. The absence of $PT$ and $T\tau$ symmetry generates an anisotropic spin splitting. \color{black}\textbf{b} The electronic bands are non-degenerate in two paths in the reciprocal space \color{black} (top diagram) \color{black} $\Gamma-H$ and $\Gamma-KM_2$.}\label{Fig:Bands}
\end{figure}

The altermagnetic nature of hematite has its origins in its crystal structure, where oxygen atoms in non-centrosymmetric positions modulated the spin densities into g-wave altermagnetic order \cite{LiborCHE,hoyer_altermagnetic_2025}. These atoms preserve the atomic rhombohedral space group $D^6_{3d}$ of hematite. In the presence of collinear altermagnetism and spin-orbit coupling, the magnetic symmetry depends on the spatial orientation of the spin order (Néel vector). The three possible magnetic point groups correspond to three possible states, which are shown in Table \ref{table:teo}.  Figure \ref{Fig:Bands}\textbf{a} shows the crystal structure of hematite in its conventional hexagonal unit cell, where the non-centrosymmetric oxygen atoms reside in the corners of the octahedra. \color{black} While inversion symmetry ($P$) alone is preserved in hematite due to its centrosymmetric crystal structure, the combinations with time-reversal ($T$) symmetry, $PT$, and translation ($\tau$) symmetry, $T\tau$, are broken. Specifically, applying time-reversal $T$ flips the spins, creating a configuration that cannot be restored to the original state by a simple translation. The subtlety of this symmetry breaking becomes evident only when explicitly considering the positions of the oxygen atoms. \color{black} The oxygen octahedrons and the collinear spin order break time-reversal and rotation $C_6$ symmetry while preserving the real-space rotation combined with spin rotation $[C_2||C_6]$, resulting in a band spin splitting, a hallmark of altermagnetic behavior \cite{PRXLibor2022}. As shown in Figure \ref{Fig:Bands}\textbf{b}, the electronic band structure reveals non-degenerate states along specific reciprocal space paths, namely $\Gamma-H$ and $\Gamma-KM_2$, further emphasizing the unique altermagnetic symmetry properties of hematite.

Our experimental approach consists of electrical transport measurements performed on C-cut-oriented Ti-doped hematite thin films ($T_\text{M} =330$ K) and XPEEM imaging on R-cut-oriented samples. This dual configuration allows us to probe all components of the anomalous Hall conductivity (AHC) $(\sigma_\text{xy},\sigma_\text{xy},\sigma_\text{xy})$ within the altermagnetic framework. 
\subsection{Electrical probe of the altermagnetic Hall effect}
\begin{figure}
    \centering
    \includegraphics[width=1\textwidth]{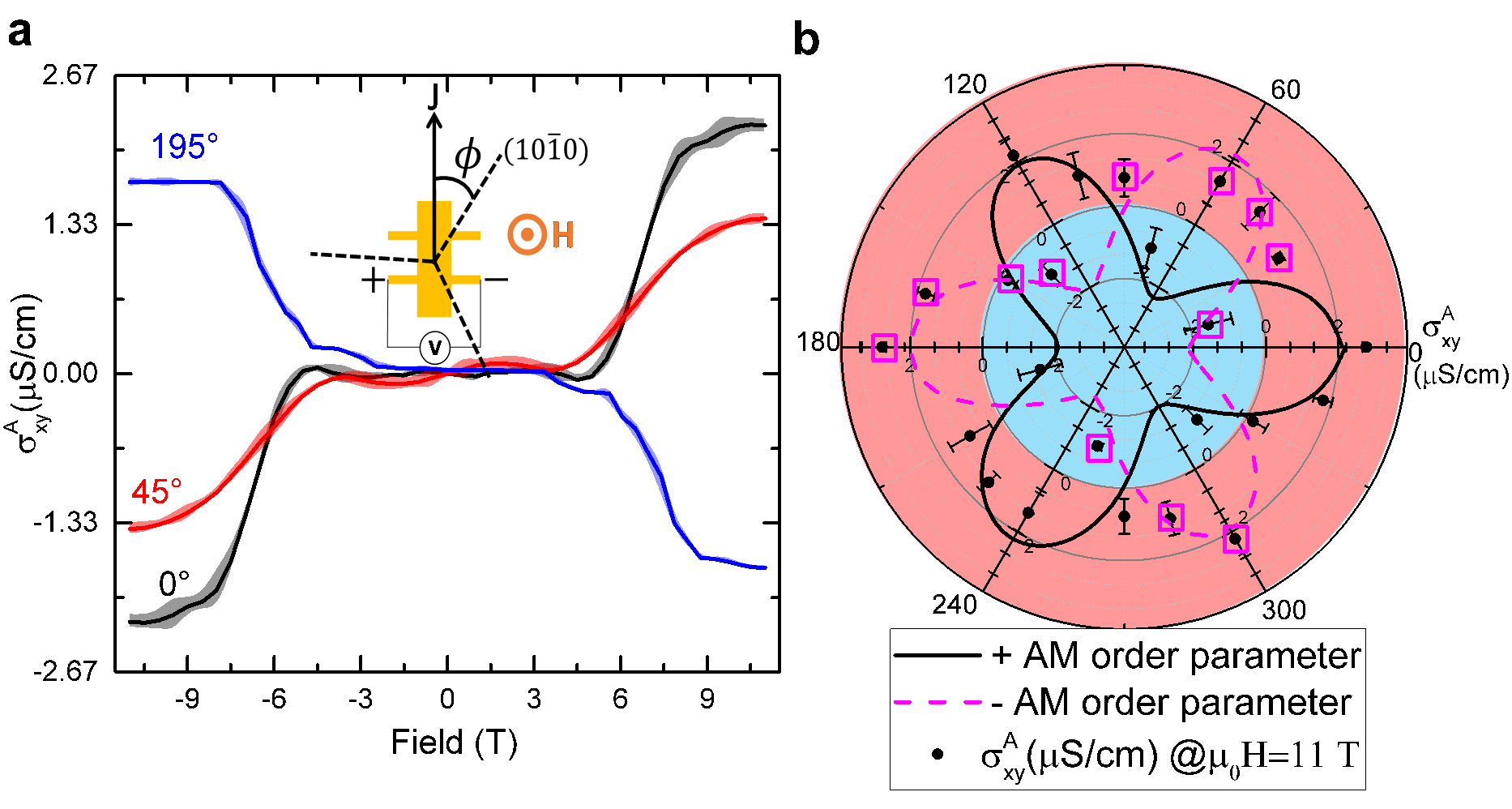}
    \caption{\textbf{a} Odd component of the transverse conductivity \color{black} in the doped sample \color{black} at 300 K of three different devices fabricated at selected angles relative to the in-plane crystallographic \textbf{a}-axis, denoted as $(10\bar{1}0)$. The shadow color indicates the standard deviation based on 20 field cycles. A sketch illustrating a device with relative orientation $\alpha$ is provided as an inset. \textbf{b} The amplitude of the AHC ($\sigma^A_\text{xy}$) \color{black} for 20 differently oriented Hall bars under an OOP field of 11 T. Two altermagnetic order parameters (magenta and black lines) of 6 nodes are plotted as a guide to the eye. \color{black}}
    \label{fig:OOP}
\end{figure}

In the (0001) plane, the measurements primarily reflect the out-of-plane $\sigma_\text{xy}$ component of the AHE, as the Néel vector rotates in the basal plane. Hall measurements are performed in Hall bars with different orientations (relative to the crystallographic \textbf{a}-axis) in the same doped-thin film. With an external out-of-plane (OOP) magnetic field, along (0001), an increase in the transverse signal is observed around 6 T at room temperature (easy axis phase, since the Morin transition is shifted to higher temperatures - see Fig. S10\textbf{b} in the Supporting Information \cite{supplemental} for details). Notably, the AHC at zero field is forbidden in the altermagnetic phase of hematite because spins are (anti)aligned along a high-symmetry direction. Breaking this symmetry by a high magnetic field induces a spin-flop, leading to a canted antiferromagnetic spin alignment with anomalous Hall components allowed. We find a large increase in our Hall signals in proximity to the spin-flop transition (Figure \ref{fig:OOP}). Above the spin-flop field, the large external magnetic field causes a change of the symmetry point group from $(R-3c, -31m)$ to $(C'_2/c', 2'/m')$. Here, the canted moment lies in the \textbf{a}-\textbf{c} plane and points towards the external magnetic field direction \cite{6-fold} \color{black}(planes in diagrams in Fig. \ref{Fig:Bands}\textbf{a}). \color{black}

The Hall signal in the collinear-oriented sublattice phase exhibits minimal response at low magnetic fields but undergoes a pronounced anomalous change near the spin-flop transition. The critical field, where the transverse resistivity starts increasing, energetically corresponds to the Morin transition in terms of magnetism. The temperature dependence of this critical field follows a square root relationship, yielding a calculated critical temperature $T_\text{M}$ of $337 \pm 8$ K (see Fig. S10\textbf{c} in the Supporting Information for details \cite{supplemental}). This behavior is consistent with the characteristics of a typical second-order phase transition, as described by Landau theory, and is in line with previous studies \cite{SMR34, SMR36}. At fields beyond the spin-flop transition, in the in-plane spin configuration, the symmetry allows for detecting the Hall pseudovector (see Table \ref{table:teo}), enabling further exploration of its properties.

Intriguing phenomena emerge when the Hall bar is patterned at orientations deviating from the high-symmetry \textbf{a}-axis, as illustrated in Figure \ref{fig:OOP}. In these configurations, we observe a transverse signal that strongly depends on the current direction and does not exhibit purely odd behavior with the magnetic field (see Fig. S2, Fig. S4\textbf{a} and Fig. S5 in the Supporting Information for the even component \cite{supplemental}). \color{black}The even-in-field signal exhibits a clear 4-fold symmetry, which is consistent with a longitudinal AMR contribution leaking into the transverse channel due to slight misalignment of the Hall bars. These contributions appear most strongly away from high-symmetry directions (0° and 90°) and are unrelated to the AHC. Their sign depends on the misalignment direction. Since the zero-field longitudinal resistance is isotropic, these effects cannot stem from intrinsic anisotropy in longitudinal conduction. Additionally, we exclude contributions from the planar Hall effect (PHE), as a $\beta$-scan preceded each measurement to confirm the out-of-plane alignment of the magnetisation to within $\pm$0.5°.  \color{black} Similar effects are observed in the high-temperature regime, though with more complex signatures, as detailed in Fig. S11 in the Supporting Information \cite{supplemental}. \color{black} While the in-plane anisotropy of hematite may play a role in shaping the AMR, the associated longitudinal (field-even) responses lie outside the scope of this study, which focuses on symmetry-driven transverse (field-odd) signals associated with the AHC. \color{black}

Remarkably, the Hall-like signals differ strongly between Hall bars oriented differently, despite the same canted moment contribution, which is fixed by the crystal orientation and the external magnetic field direction. Since the in-plane resistivity tensor of a hexagonal crystal is isotropic, these differences must arise from the spin structure \color{red} and only exist if in-plane components of the conductivity tensor are nonzero\color{black}. While the in-plane magnetic anisotropy can still play a role, the OOP magnetic field restricts the canted magnetic moment to the \textbf{a}$-$\textbf{c} plane \cite{6-fold} and therefore cannot explain the observed angular dependence.

The odd component of the transverse voltage, associated with the Hall conductivity, primarily exhibits modulation amplitudes that follow a three-fold symmetry, potentially arising from hematite's hexagonal in-plane magnetic anisotropy \cite{6-fold}. However, a key anomaly emerges: the odd signal undergoes a sign inversion at certain angles, also observed every 60° when the current direction deviates from high-symmetry axes. This behavior preserves the crystal symmetry and competes with the sixfold in-plane anisotropies of hematite. However, such sign inversion is unexpected from interactions with a positively defined energy landscape of sixfold symmetry. Importantly, crystallographic equivalent orientations exhibit comparable Hall signal amplitudes (e.g., 0° and 120°, 60° and 300°, or 15° and 135°), ruling out external influences such as differences in device patterning as the cause of these observations. \color{black} Due to the symmetry of the Hall bar, 180° differently oriented Hall bars are identical and, in consequence, also their associated transverse voltage.\color{blue} From there, a second three-fold symmetry arises (see Fig. S1\textbf{a} in the Supporting Information \cite{supplemental}).\color{black}
\subsection{Theoretical calculations of the altermagnetic anomalous Hall effect}
To understand this surprising angular dependence of the Hall signal, we carry out a theoretical analysis. To select the appropriate theoretical approach, we first determine the conduction regime. The regime is identified through temperature-dependent resistivity measurements for our samples, which were characterized by relatively low conductivity. In a semiconductor-like model (see Fig. S10\textbf{c} in the Supporting Information \cite{supplemental}), the derived activation energy, $E_\text{a}=2543 \cdot k_\text{B}$ K does not match the known band-gap of hematite. Moreover, the thermal activation factor at room temperature ($\exp(-E_\text{a}/k_\text{B}T)\approx5\times10^{-3}$) strongly indicates that the dominant transport mechanism is hopping-like transport. Given the 1\% Ti doping in our samples, a significant number of charge carriers are expected. For Ti, $n$-doping should yield a finite Hall coefficient under a single-band model, corresponding to a transverse Hall resistivity of approximately -25 $\mathrm{cm}^3\mathrm{C}^{-1}$ at room temperature. However, we observe a conventional Hall coefficient of approximately $-5\ \mathrm{cm}^3\mathrm{C}^{-1}$ over a wide temperature range, consistent with previous reports \cite{12Si, 13Si}. Additionally, the Hall voltage measured in the low-field regime matches this order of magnitude. However, the mean free path for this carrier density in the band conduction regime corresponds to an unphysically small value of 10$^{-12}$ m, further corroborating the scenario of hopping conduction within the impurity "band" of the donors with a Hall mobility of 0.1 cm$^2$V$^{-1}$s$^{-1}$. 

\begin{figure}[h]
\centering
\includegraphics[width=0.9\textwidth]{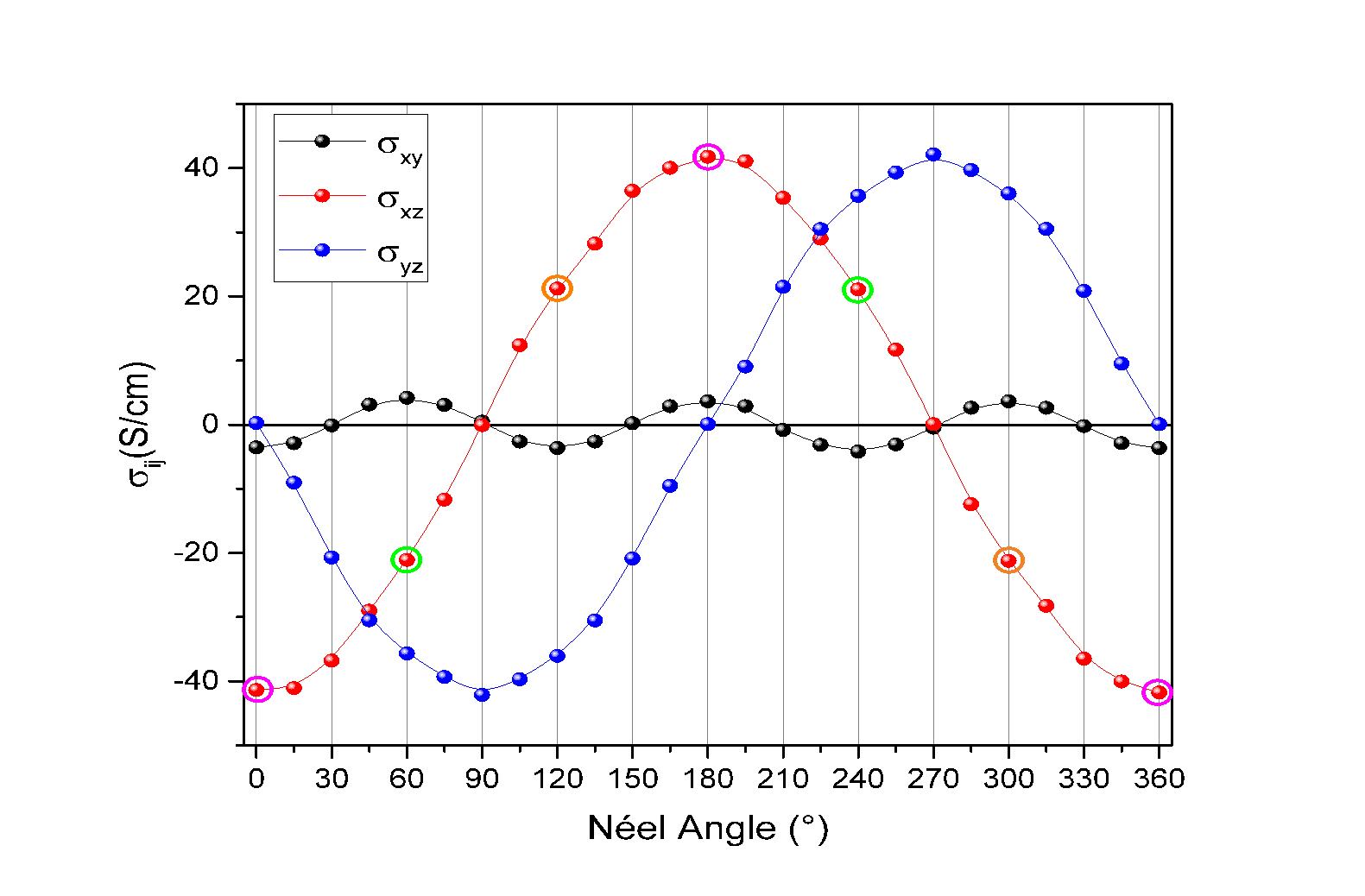}
\caption{Calculated AHC vector components $(\sigma_\text{yz},\sigma_\text{xz},\sigma_\text{xy})$ 
 \color{black} for $E-E_F=-0.12$ eV \color{black} as a function of the in-plane orientation of the Néel vector. A characteristic 180° inversion is observed with the Néel vector (color-emphasized data points), while the x and y components of the conductivity vector present a 90° shift.}\label{fig:AHC_angle}
\end{figure}

Due to the hopping-dominated transport, direct conversion of the measured transverse resistivities to Hall conductivities is not quantitatively comparable with spin-polarized conductivities derived from band structure calculations (Figure \ref{fig:AHC_angle}) \cite{PRXLibor2022, LiborCHE}. However, our transport experiments' directional dependence provides insight into the system's symmetry-driven behavior. Since the band structure is typically solved in reciprocal space (the Fourier transform of real space), the hopping process in real space can be expected to mimic the symmetry of the band structure as the same point group symmetries govern it. In the altermagnetic phase of hematite, the AHE is symmetry-forbidden because spins are (anti)aligned along a high-symmetry direction. However, we observe a notable increase in the Hall signals (Figure \ref{fig:OOP}), where anomalous Hall components become allowed due to the breaking of this symmetry.

In the high-temperature weak ferromagnetic phase, the transverse resistivity signal is in line with the expected AHE driven by a combination of the canted moment and the altermagnetic order. However, in the collinear phase, the transverse conductivity signals emerge only when the magnetic point group symmetry permits an OOP Hall pseudovector \cite{Anomalus_antiferro}. To further investigate the relationship between the low-symmetry crystal structure and the Hall signal, we analyze the dependence of the transverse resistivity on the relative orientation of the current (patterned Hall bar) and crystallographic axes. To keep the results comparable between different Hall bars, it is key that each Hall bar defines a local coordinate system, with the injected current, transverse voltage, and external magnetic field, such that positive voltage signifies conduction resembling a p-type semiconductor.

A fully antisymmetric signal in the transverse conductivity is observed when the current flows along the crystallographic axis \textbf{a}. These signals can be understood as resulting from symmetry breaking due to the reorientation of magnetic moments induced by the field above the spin-flop transition \cite{Sihematite}. \color{black} Under a large magnetic field, the sample is in a monodomain state \cite{wittmann_role_2022}, while 180° domains could exist. However, after a $\beta$-scan, a Néel vector is preferred, and there is no AHE suppression from opposite AHC. This is also observed in the sharp switching when an in-plane (IP) magnetic field is applied to invert the Néel ordering (see Fig. S3 in the Supporting Information \cite{supplemental}). Different trajectories of the spins in the spin reorientation transition can contribute to the signal as minor hysteresis (below the error bars) \cite{training, ZnHematite}. In contrast, domain sizes can be influenced by strain-induced anisotropy introduced during the sample patterning process \cite{Meer, Amin2024}, which could lead to a preferred orientation of the Néel vector in the Hall bars; such anisotropy energy is also dominated under the strong magnetic field. Because the magnetic-anisotropy energy $K_6(1-\cos{6\phi})$ is a positively defined scalar, it can modulate the magnitude but can never reverse the sign of any response.
Therefore, the 60° sign alternation can only originate from an axial Hall vector that changes sign. \color{black}

The observed crystal symmetry, combined with the magnetic order parameter, results thus in a nondiagonal conductivity tensor. The directional sign changes in the Hall signal imply that an order parameter in the system undergoes sign inversion depending on the current direction. Due to the underlying crystal symmetry, this order parameter must possess either six or twelve nodes, with maxima or nodes aligned along the crystallographic \textbf{a}-axis. This characteristic is in line with the predicted order parameter of the altermagnetic phase of hematite \cite{PRXLibor2022}. Anisotropic transport in metallic systems is often linked to the crystallographic orientation through crystal field and spin-orbit interactions \cite{Ebert_2011}. In the case of hopping-like transport, spin-dependent hopping processes in low-conductivity ferromagnets have been previously demonstrated \cite{spin-dependent-1, spin-dependent-2}, but this effect is not allowed in traditional collinear antiferromagnets. Instead, the observed Hall signal behavior serves as a hallmark of altermagnetic materials \cite{LiborCHE}. 

The altermagnetic materials allow for a local magnetization density difference between sublattices, which can explain the observed Hall sign inversion. This inversion arises from reversing local crystal anisotropies (e.g., by altering the orientation of the Hall bar) while maintaining fixed sublattice moment directions and the Néel vector \cite{Anomalus_antiferro}. In our measurement scheme, an altermagnetic spin-splitting in the x$-$y plane would be necessary to produce a spin-dependent $\sigma_\text{xy}$. A band spin-splitting in hematite is indeed observed along the $\mathrm{\Gamma-KM_2}$ directions with a resulting conductivity with three-fold symmetry, the same as our angular results. While we can thus explain our results qualitatively, quantifying the specific contribution of altermagnetic properties to the measured conductivity remains challenging within the hopping transport regime, and such investigations will have to be reserved for future work. The qualitative comparison is shown in Figure 3, where the amplitude of our result has the same angular dependence as the calculated $\sigma_\text{xy}$ (Figure \ref{fig:AHC_angle}), as the solid line shown in Figure \ref{fig:OOP}.b.

In our calculated AHC, the in-plane components have a larger amplitude compared with the OOP $\sigma_\text{xy}$ \color{red}(which depends on the Energy $E-E_f=-0.12 eV$, in our calculations)\color{black}. However, here, extinctions of the Hall conductivity at specific angles are observed in all the coefficients. \color{red} The observed sign inversions and symmetry patterns in the AHE (Fig. \ref{fig:OOP}) are a direct consequence of the vectorial structure of the AHC observed in Figure \ref{fig:AHC_angle}. \color{black} Also, our calculations show that all AHC components show a sign inversion whenever the Néel vector rotates 180°. To probe the components, which we cannot access by transport measurements, we next perform XPEEM imaging. We employ oriented samples such that a finite projection of those larger components is present for the accessible X-ray directions to obtain detectable contrast for these larger Hall components.
\subsection{Direct imaging of the altermagnetic domain spin structures}
XPEEM imaging provides information on the spin structure, with XMLD providing information on the compensated spin configuration, and the XMCD originates in the anomalous Hall pseudovector allowed within the altermagnetic framework \cite{Amin2024}. Notably, the contrast in the XMCD depends on the relative projection of each absolute amplitude of the Hall pseudovector components into the incoming X-ray direction. This imaging allows us to check the expected behavior of Figure \ref{fig:AHC_angle}, which depicts the predicted angular dependence of the AHC components.

\begin{figure}[h]
\centering
\includegraphics[width=1\textwidth]{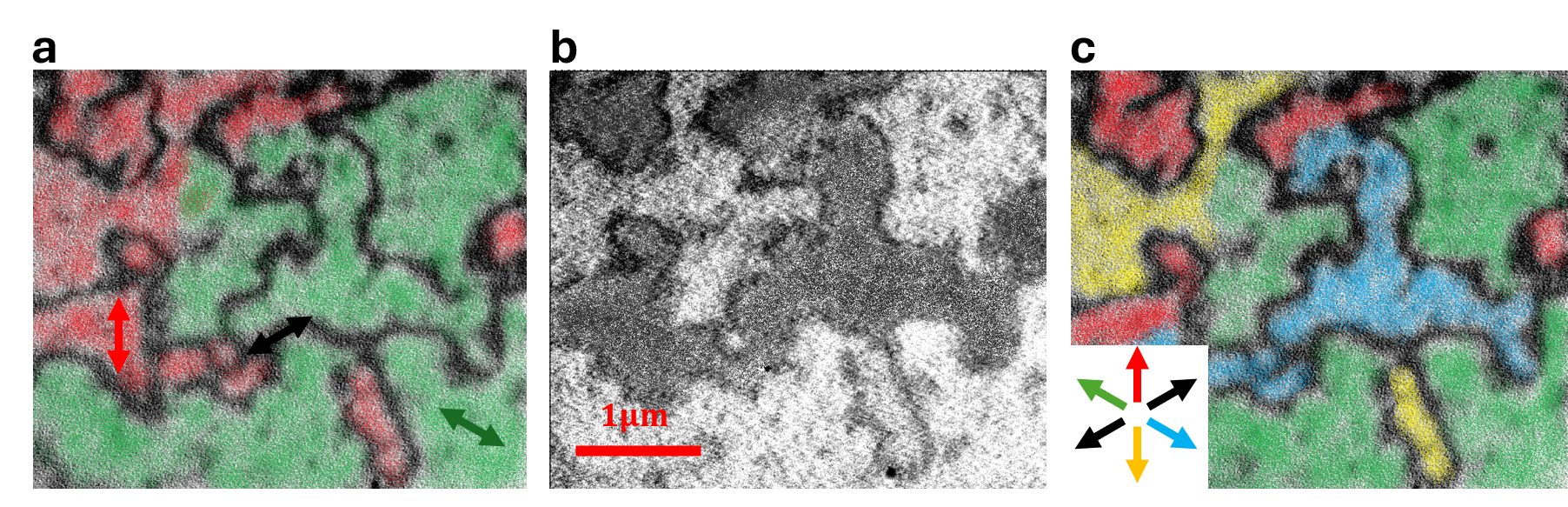}
\caption{XPEEM micrographs of \color{black} undoped \color{black}hematite in the \color{black} high temperature phase \color{black} at the $L_3$ Fe edge and room temperature. \textbf{a} XMLD-PEEM Néel vector mapping obtained by rotating the polarization plane of the incoming x-ray. \textbf{b} XMCD-PEEM, where opposite contrast levels are observed for the same Néel vector orientation, and \textbf{c} the boolean combination of linear and circular dichroism, allowing a full Néel vector mapping of the magnetic domains in hematite.}\label{XPEEM}
\end{figure}

We use XPEEM imaging to visualize the absolute Néel vector's domain structure. As shown in Figure \ref{XPEEM}, the XMLD component reveals in-plane Néel vector mapping, while the XMCD contrast probes the time-reversal symmetry breaking enabled by the altermagnetism. The presence of circular dichroism in a compensated magnet provides an experimental demonstration of the presence of the altermagnetic anomalous Hall pseudovector. A boolean combination of these imaging modes allows for a full visualization of the Néel vector in real space, only possible due to the interplay between the altermagnetic symmetry and the Hall conductivity. Our results show the power of XPEEM in probing the unique transport phenomena arising from altermagnetic symmetry, bridging real-space imaging with reciprocal-space transport measurements.

The XPEEM imaging (Figure \ref{XPEEM}\textbf{a}) reveals the intricate magnetic domain structure of hematite in line with previous observations of hematite in the easy plane phase \cite{SMRthinfilms, training, nature590}. Using XMLD-PEEM imaging (Figure \ref{XPEEM}\textbf{a}), where the intensity contrast reflects the in-plane Néel vector orientation, 180° domains exhibit identical contrast levels. The red and green regions correspond to distinct Néel vector directions, reconstructed from the dichroic contrast obtained by rotating the linear polarization of the incident X-rays. \color{black}  A third XMLD contrast is visible as narrow black stripes; however, due to their limited spatial extent and very narrow width, a reliable XMCD fitting for these regions was not possible. Consequently, the XMCD-PEEM map (Fig. \ref{XPEEM}\textbf{c}) shows only five distinct colour contrasts corresponding to the primary, more spatially extended, Néel vector orientations. \color{black}By incorporating XMCD-PEEM imaging and applying a boolean combination, unique contrast levels emerge, corresponding to the primary in-plane Néel vector orientations. The 180° flipping of the Néel vector across neighboring domains is directly associated with the predicted sign inversion of the AHE in altermagnets. These results experimentally demonstrate the connection between the altermagnetic real-space domain structure and the altermagnetic transport properties of hematite, showing an angular dependence of the AHC with the Néel vector orientation and exhibiting the altermagnetic nature of hematite.

\section{Conclusion}
We find that in the predicted altermagnet hematite, we can ascertain distinct components of the conductivity tensor by transport. The components of the Hall signal depend critically on the system's symmetry, particularly when the current direction vector deviates from high-symmetry axes. Crucially, the observed orientation-dependent behavior is independent of the constant canted moments induced during spin reorientation, which remain fixed for all device orientations at a given field strength. This highlights the symmetry-induced origin of the signal, which is in line with hematite's altermagnetic nature. To ascertain components of the anomalous Hall pseudovector not accessible by transport, we employ XPEEM imaging. 

Our XPEEM results directly visualize the altermagnetic spin structure of hematite, which is in line with the symmetry-driven transport phenomena. Specifically, the angular dependence of the Hall conductivity reveals two distinct contributions: one conventional, arising from the canted moment, and another arising from altermagnetic symmetry. We find no significant XMCD signal in the low-temperature collinear phase, which further supports the altermagnetic origin, as the electrical Hall signal is found to also only emerge above at the spin-flop transition. Notably, the observed AHE sign inversion cannot be attributed to the canted moment but rather to the altermagnetic symmetry of hematite. These measurements confirm the expected symmetry nodes of a g-wave altermagnet \cite{PRXLibor2022, LiborCHE}. Our findings show that even in the hopping regime, the altermagnetic symmetry governs the transport properties, enabling the experimental detection of anisotropic altermagnetic behavior.

So overall, by combining XPEEM imaging and angular-resolved electrical transport measurements, we demonstrate the altermagnetic nature of hematite. The appearance of the XMCD signal in the easy-plane phase and the rise of the AHE at the spin-flop transition highlight the connection between the altermagnetic order and the observed transport properties. Furthermore, the complex angular dependence of the Hall signal, including sign inversions in the odd Hall conductivity, provides direct evidence of anisotropic altermagnetic transport phenomena.

We explore a naturally insulating altermagnet that, by doping, can be rendered conducting and thus be an asset for future spintronics applications. While our results demonstrate the altermagnetic nature of hematite, to functionalize the properties, other \color{red} crystal orientations, \color{black} dopants, and doping levels can be explored in the future to tune the conductivity and transport regimes. 

\section{Experimental Section and methods}
\subsection{Sample fabrication and characterization}
High-quality (0001)-oriented films of titanium-doped hematite ($\alpha$-Fe$_{1.99}$Ti$_{0.01}$O$_3$: (Ti)$\alpha$-\ch{Fe2O3}) and pure (-1012)-hematite ($\alpha$-\ch{Fe2O3}) were grown on \textbf{c} plane sapphire and R-cut sapphire substrates, respectively, using pulsed laser deposition. Hall bars were patterned with precision at various in-plane orientations relative to the \textbf{a}-axis for the (0001)-oriented films. Detailed fabrication procedures and assessments of sample quality, including structural and magnetic characterization, are provided in the Supporting Information \cite{supplemental}, along with relevant references \cite{filmspreparation, geoHall, Zn22, Zn18, ZnHematite, lebrun2018, supl7training, 18angularc, PRXLibor2022, LiborCHE, MnTe}. The magnetic properties and quality of the pure R-cut hematite films have been independently confirmed in previous studies \cite{RcutReference}, demonstrating consistency with bulk-like hematite behavior. The films were grown using 20,000 pulses to achieve the desired thickness of approximately 150 nm, as described in \cite{filmspreparation}. The crystallographic purity of the thin films was confirmed using X-ray diffraction ({XRD}). Additionally, the in-plane crystallographic orientation of the sample was determined (see Fig. S9\textbf{c} in the Supporting Information \cite{supplemental}). 

The temperature range for the collinear phase of the hematite films was determined using superconducting quantum interference device ({SQUID}) magnetometry measurements (see Fig. S10\textbf{b} in the Supporting Information \cite{supplemental}). All the structures were fabricated using two electron beam lithography steps. First, a protective resist mask allowed ion-beam etching until the substrate was reached. Then, contacts were implemented using a lift-off technique with a Cr/Au bilayer. Different orientations were patterned in the samples, with angles of  $\phi=0,15,30,45,60,75,90,120,135,150,180,195,240,255^{\circ}$  in the plane, where $\phi=0, 120$ and $240^{\circ}$ points along one of the \textbf{a}-axes.  The devices along symmetrically equivalent crystallographic orientations are distinguished using gold markers on one of the edges of the sample, marking this orientation as the "$0^{\circ}$." Still, this choice of the origin axis is arbitrary and crystal-equivalent to the other two orientations. 

For the doped samples, magnetization measurements reveal a clear reduction in the magnetic moment below 338 K, corresponding to the Morin transition. Compared to single crystals, this transition appears broader, likely due to a multidomain state within our samples, as described by Morin et al. \cite{MorinPaper} and further discussed in our previous studies \cite{lebrun2018, training}.

In the easy-axis phase, the magnetic moment is found to be negligible, consistent with hematite's intrinsic magnetic properties. The magnitude of the canted moment in the weak ferromagnetic phase closely matches bulk values, as reported by Lebrun et al. \cite{lebrun2018} and Danneger et al. \cite{TD_Uli}. The absence of secondary phases, well-defined in-plane orientations, the observation of the Morin transition, and the agreement of the canted magnetic moment with established bulk values collectively indicate that the 1\% doping level minimally affects hematite's structural and magnetic properties.

\subsection{Electrical measurements}
The longitudinal resistance of the Hall bar was measured using a high-impedance electrometer while applying a current density of  $j=2\times10^{6}\ \mathrm{A}/\mathrm{m}^2$. A highly sensitive nanovoltmeter was used to measure the transverse voltage ($V_\text{xy}$).  Signals that are even in the current, i.e., thermoelectric voltages, Nernst effect voltage, and Righi-Leduc voltage, are one order of magnitude lower than the odd components. \color{black}The longitudinal resistivity $\rho_{xx}$ at zero field is isotropic, i.e., no variation higher than 1\% was found among the different Hall bars. To avoid contributions from field components in the plane (PHE) before every measurement, a $\beta$-scan at 11T was performed to determine the OOP direction with a precision better than 0.5° (A $\beta$-scan is an angular scan with the field perpendicular to the current direction, from IP to OOP direction). The IP projection of the $\beta$-scan helps the sample to be in a preferable monodomain state \color{black}.
In this local coordinate system, a positive transverse voltage would be detected for a p-type semiconductor, a magnetic field along the +z direction, and a current along the local +x direction. 
  
The Hall resistivity ($\rho_\text{xy}^\text{Hall}$) as a function of the external magnetic field $H$ was calculated by multiplying by the geometrical factor of the Hall bar \cite{geoHall}:
\begin{equation} \rho_\text{xy}^\text{Hall} = R_\text{xy}^\text{Hall} \cdot t \end{equation}

The longitudinal resistivity is given by:

\begin{equation} \rho_\text{xx} = \frac{R_\text{xx} \cdot w \cdot t}{a} \end{equation}

In our case, $t=150$ nm represents the thickness of the sample, $w=10\ \mu$m is the width of the Hall bar, and $a=80\ \mu$m is the distance between the two channels of the 6-point Hall bar.
 The behavior of the longitudinal resistance as a function of temperature contains key information on the transport regime of our samples. While a semiconductor model fits well, the activation energy does not correspond to the electronic gap of our material. Therefore, we consider a variable hopping regime under the arguments explained in the main text.
 
The Hall conductivity can be calculated using the values of $\rho_\text{xx}$ and $\rho_\text{xy}$:

\begin{equation}
    \sigma_\text{xy}=\frac{\rho_\text{xy}}{\rho^2_\text{xx}+\rho^2_\text{xy}}
\end{equation}
 To extract the Hall-like signal from other contributions, the transverse contribution (in our local coordinated system)  is symmetrized/antisymmetrized with respect to the external magnetic field using the equation:

\begin{equation} 2\sigma^{A(S)}_\text{xy} = \sigma_\text{xy}(H) - (+)\sigma_\text{xy}(-H) \label{eq:symm} \end{equation}

The symmetrization process is illustrated in Fig. S2 in the Supporting Information \cite{supplemental}.

\subsection{XPEEM imaging}

Doped and undoped samples were measured using XPEEM, and all samples were coated with 2 nm of Pt deposited by DC magnetron sputtering to avoid charging. The experiments were conducted at the MAXPEEM beamline of the MAX IV Synchrotron, utilizing a combination of XMLD and XMCD to probe the Néel vector and magnetic domain structure of hematite with nanoscale spatial resolution. XMLD imaging was performed by varying the linear polarization of the incoming x-rays, enabling mapping of the in-plane Néel vector, while XMCD imaging employed circularly polarized x-rays to detect symmetry-breaking effects. XMLD and XMCD measurements were performed at the Fe $L_3$-edge, where sensitivity to hematite's magnetic and electronic structure is maximized.

The XMLD asymmetry was calculated as XMLD$=\frac{I(E_1)-I(E_2)}{I(E_1)+I(E_2)}$, where $I$ is the measured pixel intensity at the maximum points of the magnetic linear dichroism spectrum at the Fe $L_3$-edge. Similarly, the XMCD asymmetry was determined as XMCD$=\frac{I(\mu^+)-I(\mu^-)}{I(\mu^+)+I(\mu^-)}$, where $I(\mu^\pm)$ represents the pixel intensity under opposite helicity polarizations at the Fe $L_3$-edge, specifically at the energy corresponding to the XMCD maximum.

To map the Néel vector, XMLD images were recorded with linear polarization angles ranging from $\theta=$ -90° to $\theta=$ 90° relative to the horizontal axis in steps of $\theta=$15°. The angular dependence of the XMLD signal was fitted using  sin$(2\theta+\phi)$, where the phase offset $\phi$ encodes the local Néel vector orientation. By combining XMLD maps with XMCD measurements, we were able to reconstruct the complete Néel vector configuration and magnetic domain structure across the hematite samples. Boolean operations between the XMLD and XMCD maps provide a comprehensive view of the Néel vector in real space, revealing both symmetry-breaking behavior and directional anisotropies linked to the altermagnetic phase.

\subsection{First-principles theory calculations}

Density functional theory (DFT) calculations were conducted using the projector augmented-wave (PAW) method as implemented in the Vienna Ab initio Simulation Package (VASP) \cite{kresse}. To account for electron correlation in localized states, we employed the Dudarev DFT+U approach, applying spherical corrections. The energy cutoff for the plane-wave basis set was set to 520 eV to ensure convergence. Exchange-correlation effects were treated using the Perdew–Burke–Ernzerhof (PBE) functional within the generalized gradient approximation (GGA). The Brillouin zone was sampled using a Monkhorst-Pack k-point grid of 11×11×11. Maximally localized Wannier functions (MLWFs) were constructed using the Wannier90 code \cite{Wannier} to facilitate accurate interpolation of the electronic structure. The Berry curvature formalism determined the intrinsic Hall conductivity, employing a dense Brillouin zone mesh of 240×240×240 points.

\medskip
\textbf{Supporting Information} \par 
Supporting Information is available from the Wiley Online Library or from the author.

\medskip
\textbf{Acknowledgements} \par 
All authors from Mainz acknowledge support from SPIN + X (DFG SFB TRR 173 No. 268565370, projects A01, A11, B02, and project \# 423441604). M.K. acknowledges support from the Research Council of Norway through its Centers of Excellence funding scheme, project number 262633 "QuSpin." M.K. and E.G.-R. also acknowledge the support and funding from the project CRC TRR 288 - 422213477 Elasto-Q-Mat (project A12), and the European Commission through the Horizon Europe framework program with project HORIZON-CL4-2021 \# 101070287 (SWAN-ON-CHIP). E.G-R., C.S., L.L., W.Y., and M.K. recognize the exceptional support of the MAXPEEM beamline scientists under proposal No. 20231423.  Research conducted at MAX IV, a Swedish national user facility, is supported by the Swedish Research Council under contract 2018-07152, the Swedish Governmental Agency for Innovation Systems under contract 2018-04969, and Formas under contract 2019-02496. C.-Y.Y. and E.B. acknowledge support from the National Research Foundation of Korea (Grant No. 2021M3F3A2A01037525).

\textbf{Declarations}
There is no conflict of interest or competing interests for any author.
\textbf{Contributions}
M.K., G.J., and L.S. proposed the experiment.
R.G.-H. and E.G.-R. did the theory calculations with L.S.'s help. L.S. received help with the theory calculations from V.K. and R.U.
Av. R. fabricated and optimized the doped samples.
A.A. fabricated the non-doped samples with the help of E.B.
E.G.-R. and S.D. fabricated the devices with the help of E.B. and C.S.
E.G.-R. performed the electrical measurements with the help of C.S., F.F., and S.D.
E.G.-R., C.S., L.L., and W.Y. performed XPEEM imaging in MAXIV under the supervision and help of beamline scientist E.G.
E.G.-R., C.S., F.F., L.L., and W.Y. processed the data and XPEEM images.
E.G.-R analyzed the data, with the help and discussion of C.S., F.F., L.S., J.S., G.J., and M.K.
E.F.-G. wrote the manuscript with input from C.S., R.G.-H., F.F., A.R., C.-Y. Y., G.J., and M.K. All the authors revised the manuscript.

\medskip

\bibliographystyle{unsrt}
\bibliography{ArXiv}

\end{document}